\documentclass[final,english]{bullsrsl}[2022/06/15]

\usepackage[latin1]{inputenc}
\usepackage[T1]{fontenc}
\usepackage{natbib} 
\usepackage{graphicx}

\begin{document}
\title{Initial results of our spectro-photometric monitoring of XZ Tau}

\author[affil={1},corresponding]{Arpan}{Ghosh}
\author[affil={1}]{Saurabh}{Sharma}
\author[affil={2}]{Joe Philip}{Ninan}
\author[affil={2}]{Devendra K.}{Ojha}
\author[affil={1}]{Aayushi}{Verma}
\author[affil={1}]{Tarak Chand}{Sahu}
\author[affil={3}]{Rakesh}{Pandey}
\author[affil={2}]{Koshvendra}{Singh}
\affiliation[1]{Aryabhatta Research Institute of Observational Sciences (ARIES) Manora Peak, Nainital 263 001, India}
\affiliation[2]{Department of Astronomy and Astrophysics, Tata Institute of Fundamental Research (TIFR), Mumbai 400005, Maharashtra, India}
\affiliation[3]{Physical Research Laboratory, Navrangpura, Ahmedabad - 380 009, India}
\correspondance{arpan@aries.res.in}
\date{5th June 2022}
\maketitle

% Abstract of the paper in the same language as the paper
\begin{abstract}

We present here initial results of our spectro-photometric monitoring of XZ Tau. During our  
monitoring period, XZ Tau exhibited several episodes of brightness variations in timescales of months 
at optical wavelengths in contrast to the mid-infrared wavelengths.
The color evolution of XZ Tau during this period suggest that the brightness variations are driven by
changes in accretion from the disc. The mid-infrared light curve shows an overall decline in 
brightness by $\sim$ 0.5 and 0.7 magnitude respectively in WISE W1 (3.4 $\mu$m) and W2 (4.6 $\mu$m) bands. The emission profile of 
the hydrogen recombination lines along with that of Ca II IRT lines points towards magnetospheric 
accretion of XZ Tau. We have detected P Cygni profile in H$\beta$ indicating of outflowing winds from
regions close to accretion. Forbidden transitions of oxygen are also detected, likely indicating of
jets originating around the central pre-main sequence star.

\end{abstract}


%\section{Section -- Level 1 title (Times New Roman, bold, 14 pts)}
\section{Introduction}

%Stars form from the gravitational collapse of the giant molecular clouds (GMCs). 
The process of accretion is fundamental in the formation of stars even though it is poorly understood \citep{2016ARA&A..54..135H}. Initially, a steady state accretion rate was theorized for the 
formation of stars \citep{1969MNRAS.145..271L,1977ApJ...214..488S,1984ApJ...286..529T}. 
However, the observed discrepancy in the luminosity of Class {\sc I} young stellar objects 
(YSOs) with that of the theoretical models gave rise to the `Luminosity Problem'
\citep{1990AJ.....99..869K,2009ApJS..181..321E}. The idea of `episodic accretion' at early  
stages of pre-main sequence (PMS) evolution came up as a possible solution to the luminosity 
problem. Observations indicate that episodic accretion phenomena spans the entire PMS evolutionary stages from
Class {\sc 0} to Class {\sc II} \citep{2015ApJ...800L...5S}. Various theoretical models has 
been proposed to explain the origin of the circumstellar disk instabilities that lead to the 
episodic outbursts. These models range from gravitational instabilities to external 
perturbations by an eccentric binary \citep{2014prpl.conf..387A}.

  The phenomenon of episodic accretion was first observed in 1936 when FU Ori underwent an outburst of $>$ 5 magnitudes in V band. Since then, around 33 sources have been discovered 
  that have demonstrated episodic accretion behaviour with outburst magnitudes $>$ 2 
  magnitudes in V band, along with a wide variety of rise and decay timescales \citep{2014prpl.conf..387A}. YSOs that 
  exhibit episodic accretion behaviour are classified as FUors and EXors using a binary 
  classification approach. FUor outbursts are manifested as 4-5 magnitude variation in V band
 with absorption features in their spectrum and outburst timescales of decades. EXors on the other hand, undergo outburst of 2-3
 magnitudes with their spectrum consisting of emission features and outburst timescales of 1-2 years. 

   In this paper, we will be presenting initial results of our spectro-photometric 
   monitoring of a YSO, XZ Tau. \citet{2009ApJ...693.1056L} classified XZ Tau as a bonafide EXor. It 
   is located at an approximate distance of 140 pc in the L1551 dark cloud of Taurus star-forming region. Previously, \citet{2012ApJ...749..188L,2014prpl.conf..387A} have reported irregular
   brightness variations in optical wavelengths with timescales of months. They have attributed these 
   variations to short scale enhancements of disk accretion. However, simultaneous multi-wavelength   
   photometric monitoring backed with spectroscopic observations is lacking. Such studies will enable us 
   to understand the origin of such small scale enhancements in accretion rate, thus motivating us to 
   carry out the monitoring of XZ Tau. This paper is arranged as follows : Section \ref{obs} describes
   about the observations and the data reduction techniques employed. In Section \ref{res}, we 
   describe the photometric and spectroscopic evolution of XZ Tau during our monitoring period and finally
   in Section \ref{discussion} we conclude this paper based on our findings in Section \ref{res}.

%\subsection{Subsection -- Level 2 title (Times New Roman, bold, 12 pts)}
\begin{table*}
\centering
\begin{minipage}{88mm}
\label{tab:obs_log}
\caption{Log of spectroscopic observations.}
\end{minipage}
\bigskip

\begin{tabular}{cccccc}
\hline
\small
\textbf{Telescope}  & \textbf{Date} & \textbf{Grism} & \textbf{Wavelength} & \textbf{Resolution} & \textbf{Exposure} \\
                 &  &  & \textbf{Range} & {\bf R} & \textbf{Time} \\
\hline
3.6m DOT & 2020 October 23  & Cross-dispersed  & 0.55-2.5 $\mu$m & $\sim$ 1500 & 120 sec $\times$ 4 frames\\
(TANSPEC)&                  &                  &                 &             &                           \\  
2m HCT   & 2023 February 22 & Gr7 & 0.38-0.68 $\mu$m & $\sim$ 1200 & 1500 sec \\
(HFOSC)  &                  &                  &                 &             &                           \\
2m HCT   & 2023 February 22 & Gr8 & 0.58-0.84 $\mu$m & $\sim$ 2000 & 1500 sec \\
(HFOSC)  &                  &                  &                 &             &                           \\
%abc & 123 & 456 & 789 \\
%abc & 123 & 456 & 789 \\
\hline
\end{tabular}
\end{table*}

\begin{figure}
\centering

\includegraphics[width=0.99\textwidth]{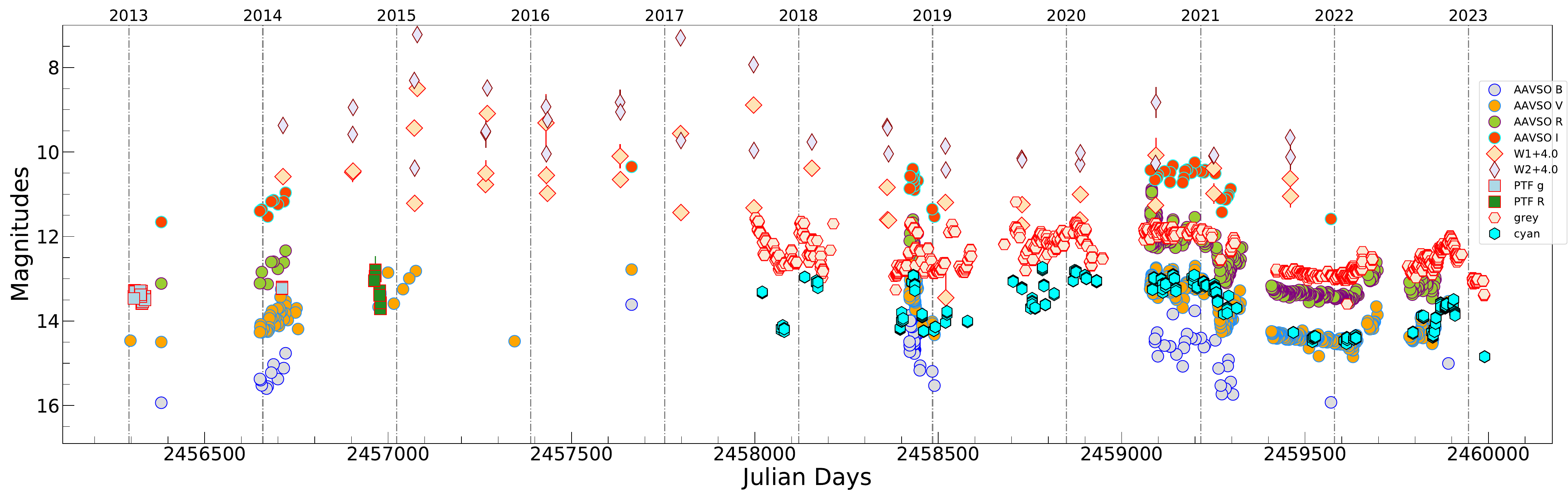}
\begin{minipage}{12cm}
\caption{Light Curve of XZ Tau in Johnson-Cousins B, V, R$_C$ and I$_C$, ZTF zg and zr, PTF g' and R and NEOWISE W1 and W2 bands. The AAVSO survey multi-band photometric data is labeled with AAVSO prefix whereas the Palomar survey data is prefixed with PTF. The ATLS survey data is denoted by the cyan and orange levels respectively. The NEOWISE W1 and W2 magnitudes are scaled by 4 magnitudes so as bring forth the variations in light curve more clearly.
}
\end{minipage}
\label{fig1}
\end{figure}

\section{Observation and Data Reduction} \label{obs}

\begin{figure}
\centering
\includegraphics[width=0.45\textwidth]{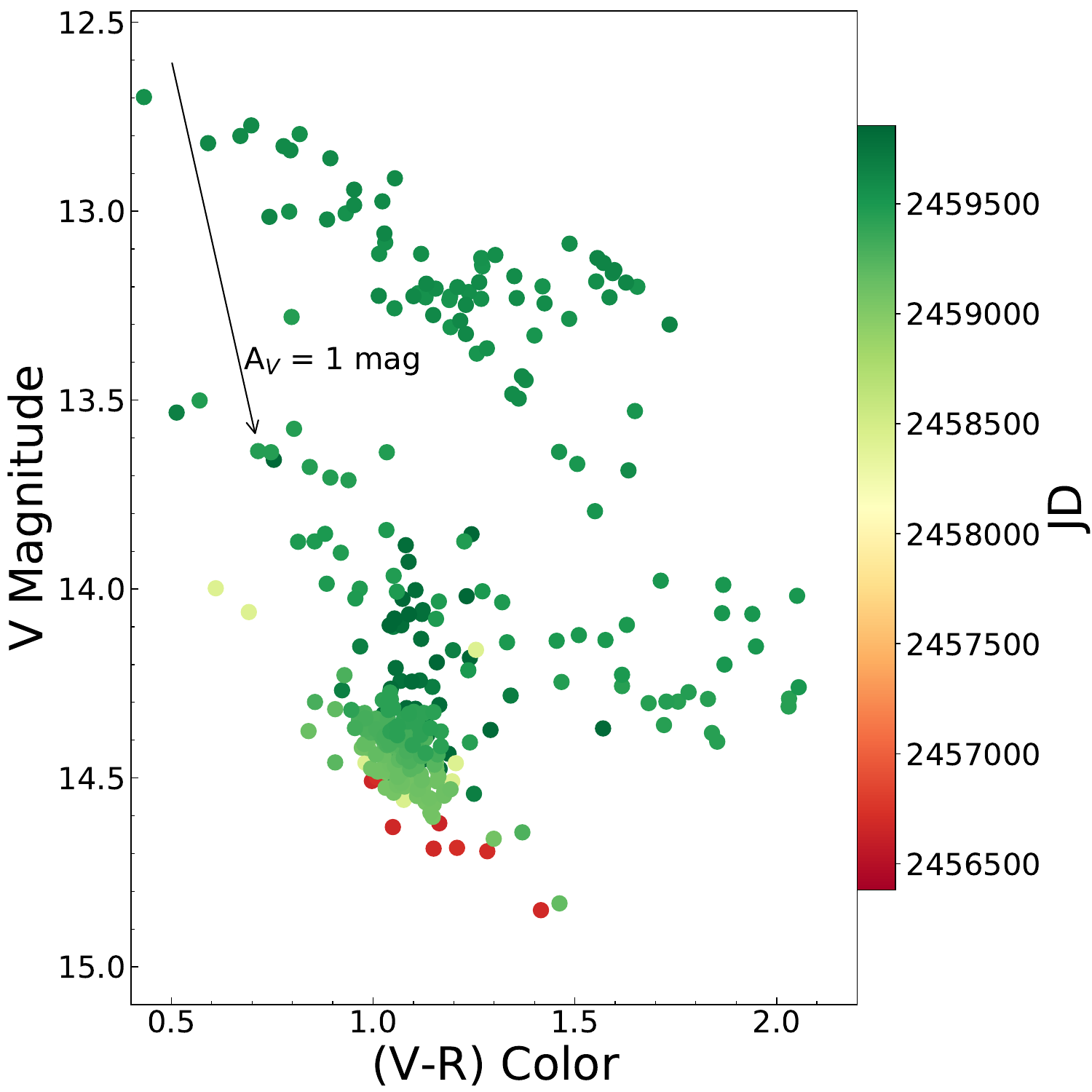}
\includegraphics[width=0.45\textwidth]{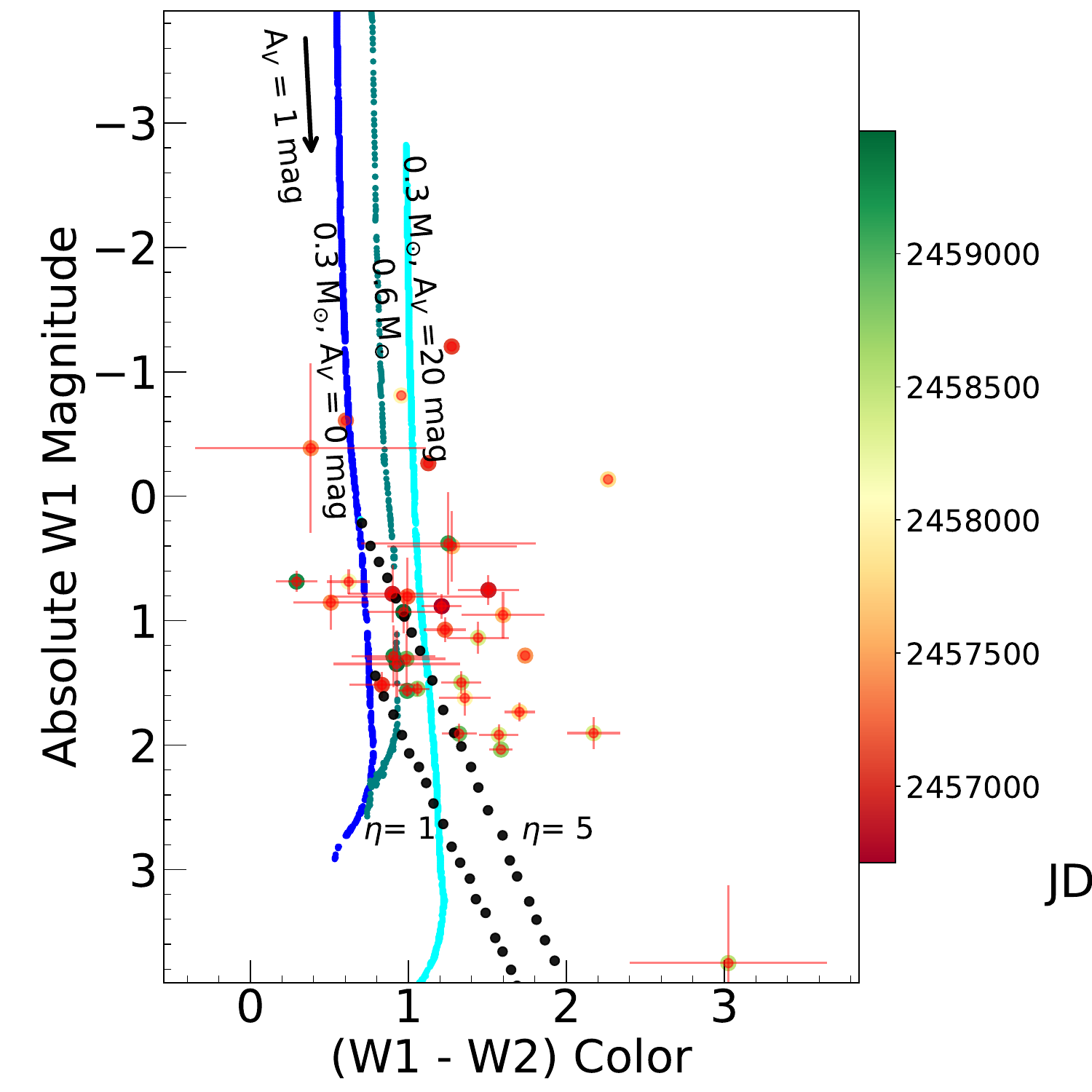}
\begin{minipage}{12cm}
\caption{Left panel shows the color evolution of XZ Tau in the V vs (V-R) color-magnitude plane between 2013 and 2023. A reddening vector of A$_V$ = 1 mag is drawn to show the photometric color evolution based on extinction alone. %{\bf The position of XZ Tau in the optical color magnitude diagram for the time period after mid-2021 is probably due to a combination of decrease in the accretion flux  and increase in extinction which resulted in the red-ward movement of XZ Tau in CMD plane.} 
The right panel shows the evolution of XZ Tau in the W1 absolute magnitude vs W1 - W2 color plane with time. The model isomass curves obtained from \citet{2022ApJ...936..152L} for 0.3 M$_{\odot}$, A$_V$ = 0 mag and A$_V$ = 20 mag and for mass 0.6 M$_{\odot}$ is plotted to highlight the movement of XZ Tau in MIR CMD across the transition ($\eta$ = 1) and sufficient dominance ($\eta$ = 5) regions due to changes in accretion. }
\end{minipage}
\label{fig2}
\end{figure}

\subsubsection{Photometric Data}
We have obtained multi-epoch photometric data of XZ Tau in broadband optical and 
mid-infrared (MIR) filters. The optical monitoring was obtained using the archival data from
Palomar Transient Factory (PTF) \citep[see details][]{2009PASP..121.1395L}, Asteroid 
Terrestrial-impact Last Alert System (ATLAS) \citep[see details][]{2018PASP..130f4505T}, 
Zwicky Transient Facility (ZTF) \citep[see details][]{2019PASP..131a8002B}, and American 
Association for the Variable Star Observers (AAVSO). The optical data was obtained 
in Johnson-Cousins (AAVSO observations are in Johnson-Cousins filter system) $B$ 
(0.44 $\mu$m), $V$ (0.55 $\mu$m), $R$ (0.65 $\mu$m) and $I$ (0.80 $\mu$m), PTF $g$ (0.477 
$\mu$m) and $R$ (0.63 $\mu$m), ZTF $zg$ (0.48 $\mu$m) and $zr$ (0.64 $\mu$m) and ATLAS $cyan$ 
(0.53 $\mu$m) and $orange$ (0.68 $\mu$m) bands. Time series MIR data was obtained from the 
Near-Earth Object Wide-field Infrared Survey Explorer (NEOWISE) survey \citep[for details, see][]{2014AAS...22321708M}. The MIR observations are in $W1$ (3.4 $\mu$m) and $W2$ (4.6 $\mu$m) 
bands and are obtained from the NASA/IPAC Infrared Science 
archive (\url{https://irsa.ipac.caltech.edu/Missions/wise.html}).

\subsubsection{Spectroscopic Data}
We have spectroscopically monitored XZ Tau in two epochs using the Hanle Faint Object 
Spectrograph Camera (HFOSC) of 2m Himalayan {\it Chandra} Telescope (HCT) (details of 
the HFOSC instrument is available at \url{https://www.iiap.res.in/iao/hfosc.html}) and 
the TIFR-ARIES Near-infrared Spectrometer (TANSPEC) of 3.6m Devasthal Optical Telescope (DOT)
\citep{2022PASP..134h5002S}. Table \ref{tab:obs_log} contains the log of spectroscopic observations. 

   The HFOSC observations were obtained using grisms Gr 7 and Gr 8 which provide a resolution
of R $\sim$ 2000 providing wavelength coverage of 0.4 to 0.9 $\mu$m. The HFOSC spectrum was reduced
and calibrated using the standard IRAF modules. IRAF is distributed by National Optical 
Astronomy Observatories, USA which is operated by the Association of Universities for
Research in Astronomy, Inc., under cooperative agreement with National Science Foundation for
performing image processing. The details of the reduction and calibration 
procedures are described in detail in \citet{2023arXiv230600759G}. TANSPEC observations were 
obtained using 1" slit that provides a resolution of R $\sim$ 1500 providing a wavelength 
coverage of 0.55 to 2.5 $\mu$m. The TANSPEC spectrum is reduced using the TANSPEC data 
reduction pipeline named pyTANSPEC \citep{2023JApA...44...30G}. The other details of the 
TANSPEC data reduction procedure has been outlined in \citet{2023arXiv230600759G}.

\section{Results and Analysis} \label{res}

\begin{figure}[h]
\centering
\includegraphics[width=0.90\textwidth]{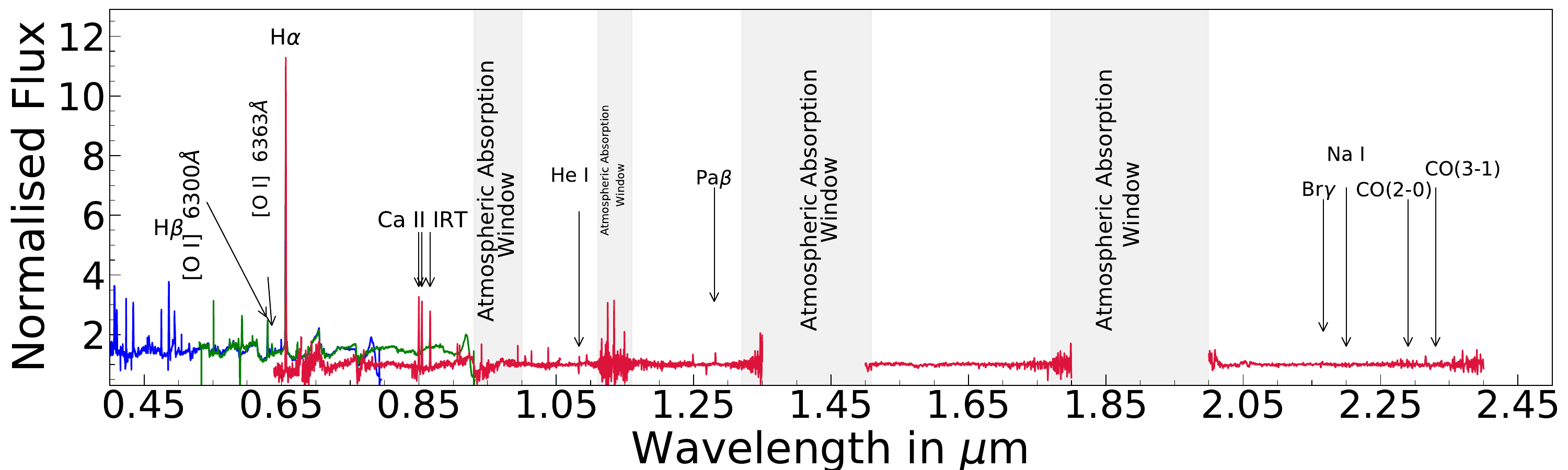}
\includegraphics[width=0.90\textwidth]{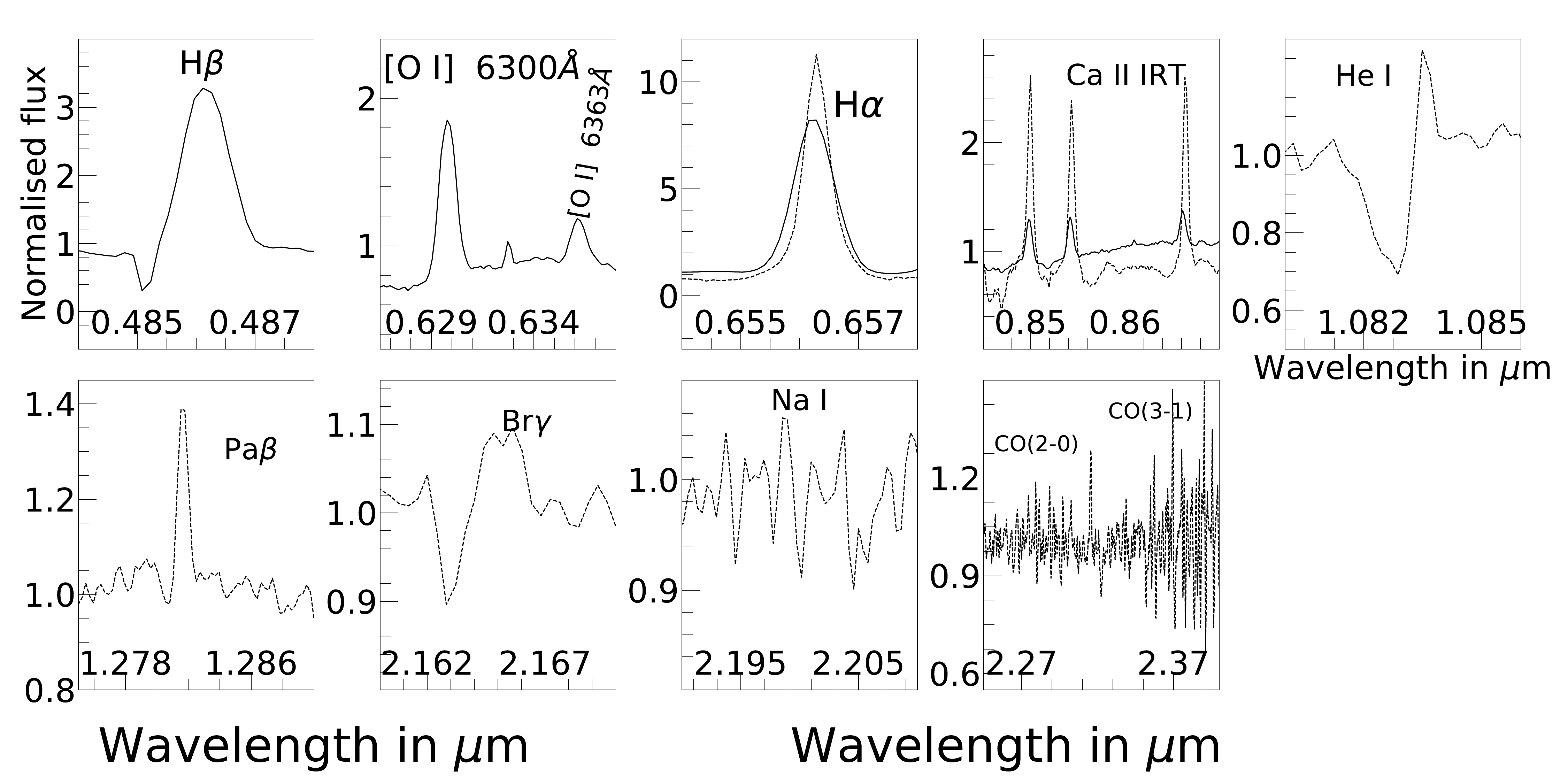}
\begin{minipage}{12cm}
\caption{Top panel shows the continuum  normalised spectra of XZ Tau obtained during our monitoring period using HFOSC with 2m HCT ( $\sim$ 0.4-0.9 $\mu$m) and TANSPEC with 3.6m DOT ($\sim$ 0.65-2.4 $\mu$m). The HFOSC spectra is taken using the Gr7 and Gr8 grisms and are denoted by blue and green colors respectively. The TANSPEC spectrum is denoted in red color. Important spectral features of XZ Tau are marked in the spectra. Bottom panel shows the zoomed in view of the line profiles that have 
been used in the present study. The HFOSC and TANSPEC line profiles are denoted by the solid and dashed lines respectively. }
\end{minipage}
\label{spectra}
\end{figure}

\subsection{Light Curve} 

Figure \ref{fig1} shows the light curve (LC) of XZ Tau spanning almost ten year period, 
beginning from 2013 to 2023. The typical uncertainties in the optical bands is below 0.05 magnitudes. During this period, XZ Tau has exhibited multiple episodes of 
brightness variations in optical bands. XZ Tau brightened by about 1.3, 1.1, 0.9, and 0.8 magnitudes in optical $B$, $V$, $R$, and $I$ bands respectively between March 28, 2013 and 
March 03, 2014.  We do not have photometric coverage in the subsequent months of 2014, 
however, we see another episode of 1 magnitude brightening in $V$ band between November 
09, 2014 and February 21, 2015. Next we observe that XZ Tau is displaying a peak brightness of 11.5 
magnitude in $orange$ band on September 03, 2017 and is gradually decaying to 12.6 magnitude on  
December 26, 2017. It started to brighten again, reaching 11.6 magnitude on January 06, 2018 after 
which it gradually faded to 13.0 magnitude in $orange$ on March 06, 2018. Between 
September 18, 2018 and November 01, 2018 XZ Tau brightened from 12.9 magnitude in $orange$ to 11.6 
magnitude. During the same time period, it brightened by $\sim$ 1 magnitude in $B$ and $V$
bands reaching peak brightness of 13.6 and 12.5 magnitudes respectively. The $R$ and $I$ bands
also display similar evolution during this period, reaching peak magnitudes of 11.5 and 10.4
respectively. It started to decay from peak brightness reaching a minimum on January 06, 2019
with $\Delta$m of $\sim$ 1.2, 2, 2 and 1 magnitude in $orange$, $B$, $V$, and $I$ bands respectively. 
Another brightening event was recorded by the ATLAS survey starting from September 10, 2019 and 
reaching maximum brightness of 11.7 and 12.8 magnitudes in $orange$ and $cyan$ bands 
respectively on January 26, 2020. During this interval, $\Delta$m was 1 magnitude in both 
$orange$ and $cyan$ bands. After January 26, 2020, XZ Tau faded by approximately 0.5 magnitudes 
in both $orange$ and $cyan$ bands. The brightness of XZ Tau remained almost constant for the 
rest of 2020. In 2021, XZ Tau exhibited a sudden fading event with $\Delta$m of $\sim$ 1 magnitude
 in $orange$, $B$, $V$ and $I$ band respectively. Following this fading event, the brightness
 of XZ Tau started to rise almost reaching the brightness levels of the pre-fading level. 
The LC of XZ Tau was stable for the rest of 2021. Starting from January 01, 2022 the $R$ band 
brightness rose by $\sim$ 0.8 magnitude to 12.6 while the variation in $\Delta V$ and 
$\Delta orange$ was $\sim$ 0.9 magnitude and 0.7 magnitude respectively, with peak $V$ and 
$orange$ band magnitudes to be 13.6 and 12.4 respectively. These brightness 
variations occurred over a period of 113 days. Another brightness variation is observed 
beginning from September 25, 2022 in which $\Delta$m varied by almost 0.8, 1 and 1 magnitudes in $orange$, $V$ and $cyan$ bands after which the magnitudes returned to their pre-brightening levels of 13.4 and 14.8 in $orange$ and $cyan$ bands on February 14, 2023.

%Small scale brightness variations were 
%again observed at the beginning of 2020 as evident from the ATLAS bands. During this episode 
%the source brightened from x,y magnitudes in orange and cyan bands respectively to x2,y2 
%magnitudes over a period of m days. XZ Tau displayed a period of almost constant brightness 
%from mid of 2020 to early 2021 after which it displayed a sudden drop in brightness. This 
%magnitude variations has been observed in both the AAVSO and ATLAS surveys. The LC of XZ Tau
%shows a plateau phase from around mid 2021 to early 2022 when it started to brighten again. %In 
%2022 XZ Tau exhibited two short scale brightening events of x3,y3 and x4,y4 respectively as 
%evident from the ATLAS surveys.

    The MIR LC of XZ Tau shows an overall decay in brightness from February 25, 2014 to the end 
    of our coverage on September 03, 2021. At MIR wavelengths, XZ Tau has faded by $\sim$ 0.5 
    magnitude in $W1$ band and 0.7 magnitude in $W2$ band respectively. The MIR LC displays 
    a huge scatter which can be attributed to the fact that individual NEOWISE measurements 
    are noisy. The uncertainty in individual NEOWISE measurements can arise from a variety of factors which are highlighted in the NEOWISE website (\url{https://wise2.ipac.caltech.edu/docs/release/neowise/expsup/sec3_2.html}). The uncertainty in individual NEOWISE measurements is below 0.05 magnitudes. Further analysis of MIR data has been done by taking the median value of the individual magnitude measurements for a given julian day (typically 3-5 exposures for a given julian day) and the error is estimated by taking the standard deviation of the magnitude variations. This has been done to estimate the error in the individual measurements as the photometric error in individual measurements are small. Previously, similar method has been employed in the case of V2493 Cyg by \citet{2023arXiv230600759G} to deduce its long term brightening from the similarly noisy individual NEOWISE measurements. 

    The brightness variations observed in the LC can be due to fluctuations in the accretion 
    rate or due to change in line of sight extinctions or a combination of both. We will investigate the variations in the next section based on the color evolution of XZ Tau.  

\subsection{Color Analysis}

In this section, we will investigate about the brightness variations that we observe in the LC of XZ 
Tau. YSOs like XZ Tau display brightness 
fluctuations that are driven by several factors. According to \citet{2018JAVSO..46...83H}, the PMS 
variability can be classified into three broad categories : Type-I variability which is attributed to
the stellar spots modulated by the stellar rotation period having typical variability amplitude of 0.1 magnitude. Type-II variability which is driven by 
accretion. Type-III variability also known as ``dippers'' occurs due to the occultation of the central PMS star by matter within the disk, that causes the
observed variations. The Type-II and Type-III variability can be distinguished by their slopes of 
color evolution in the color-magnitude diagram (CMD) plane. The dippers follow the slope of the 
extinction vector whereas 
Type-II sources exhibit distinct slope different from the extinction vector. This scheme has 
previously been followed by \citet{2022AJ....163..263H}. Left panel of Fig \ref{fig2} shows the color evolution
of XZ Tau during our monitoring period in V vs (V-R) CMD plane. We have also drawn an extinction 
vector of A$_V$ = 1 magnitude to compare the slope of the color variations with that of the extinction
variations. From the figure (left panel of Fig \ref{fig2}), it is clearly evident that the slope of the color evolution of XZ Tau is
distinctively different from that of the extinction vector. This implies that the probable cause of 
the observed variability to be driven by accretion as dominant factor. This implies that XZ Tau is 
possibly displaying variability of Type-II type. From the left panel of Fig \ref{fig2}, it also evident that there is a distinct clustering of the position of XZ Tau in the optical CMD for the time period between mid-2021 to mid 2022 which is probably due to a combination of decrease in the accretion flux and increase in extinction which resulted in the red-ward movement of XZ Tau in CMD plane.

    Recent theoretical advancements in the field of episodic accretion by \citet{2022ApJ...936..152L}
    have led us to better understanding of the MIR color evolution. Their models define regions of the MIR CMD plane into transition ($\eta$ = 1) and sufficient dominance regions  ($\eta$ = 5) which are based on the competitive dominance between the photospheric emission and disk emission. At low accretion rates, the stellar photospheric radiation is dominant with the contribution of the redder flux coming from disk and vice versa \citep{2022ApJ...936..152L}. The MIR color evolution of XZ Tau also points towards gradual decline in accretion rate resulting in migration towards the $\eta$ = 1 line from $\eta$ = 5 line as evident from the right side of Figure \ref{fig2}.

\subsection{Spectral Features}

Figure \ref{spectra} shows the continuum normalised medium resolution spectra of XZ Tau obtained using TANSPEC and HFOSC respectively. Bottom panel of Figure \ref{spectra} shows the zoomed in view of the line profiles that are presented here. The main spectroscopic features that
are identified in the TANSPEC spectrum are the absorption features in CO bandheads of (2-0) and (3-1) transitions, 2.2 $\mu$m
Na I  and 1.083 $\mu$m He I line. The spectrum also exhibits emission features 
in hydrogen recombination lines of Brackett$\gamma$, Paschen$\beta$, H$\alpha$ and Ca II infrared 
triplet (IRT) lines. A detailed analysis of the line profiles has already been reported in our 
previous work \citet{2023JApA...44...30G}. In summary, the observed line profiles with TANSPEC point
towards the magnetospheric accretion regime in XZ Tau. The HFOSC spectrum is obtained almost two and
half years after the TANSPEC spectrum. Comparing our HFOSC spectrum with that of the optical 
part of the TANSPEC spectrum, we observe that the spectrum still exhibits the emission features in 
H$\alpha$ and  Ca II IRT but the normalised fluxes of the spectral features are reduced in the latest
spectra. This is also supported by our photometric monitoring which clearly shows that the magnitude of
XZ Tau dimmed by $\sim$ 1.4 and 1.7 magnitudes respectively in $orange$ and $cyan$ bands between 2020 October and 2023 February.
The decrease in normalised fluxes of H$\alpha$ and Ca II IRT can be likely attributed to decrease in 
accretion rate. The most interesting spectroscopic features of XZ Tau that we have obtained with are 
the P Cygni profile in H$\beta$ and the forbidden transitions of oxygen at 6300 \AA~ and 6363 \AA~.
The observed P Cygni profile in H$\beta$ line is indicative of outflowing winds from regions close to 
accretion as it has similar origin to H$\alpha$ \citep{2022ApJ...926...68G}. The forbidden emission
lines in YSOs are believed to originate in jets and winds around the central star 
\citep{1997IAUS..182..443K}. The presence of the forbidden transitions in oxygen at 6300 and 6363 \AA~ therefore indicate outflowing jets from XZ Tau.

\vspace{-0.75cm}
%\subsubsubsection{Subsubsubsection -- Level 4 title (Times New Roman 12 pts)}
\section{Discussion and Conclusion} \label{discussion}

We have presented here the initial results of our spectro-photometric monitoring of XZ Tau. During 
our monitoring period, XZ Tau has exhibited multiple episodes of small scale brightening events 
of $\Delta$m $\sim$ 1 magnitude in the optical regime. Such magnitude variations were not observed 
in the MIR regime during our monitoring period. The MIR magnitudes show an overall decline by $\sim$
0.5 and 0.7 magnitudes respectively during our monitoring period. The optical color evolution of XZ 
Tau during our monitoring period can possibly be attributed to the variations in the accretion rate. 
In this regard, we can possibly classify the variability exhibited by 
XZ Tau to be of Type-II variability based on \citet{2018JAVSO..46...83H} classification scheme. We 
have monitored XZ Tau spectroscopically on two epochs. The emission features in the hydrogen 
recombination lines and the Ca II IRT lines implies that XZ Tau to be accreting via the 
magnetospheric accretion regime \citep{1997IAUS..182P.272F,1998AJ....116..455M}.  There is a decrease 
in the normalised flux of H$\alpha$ and Ca II IRT lines between our two epochs of observations which
further lend weight to the observed photometric color variations being driven by changes in accretion rate. The HFOSC spectrum of XZ Tau displays a P Cygni profile in H$\beta$, indicative of 
outflowing 
winds from regions close to accretion. Forbidden lines of oxygen [O I] $\lambda$6300 and 
$\lambda$6363 \AA~ are also observed in the HFOSC spectrum of XZ Tau which indicates the presence of 
outflowing jets from XZ Tau. We plan to further undertake spectro-photometric observations of XZ Tau
in future to understand the evolution of the accretion and outflowing jets and winds with that of the
photometric color changes. 

\vspace{-0.7cm}
\begin{acknowledgments}

We thank the staff at the 3.6m DOT, Devasthal (ARIES), for their co-operation
during observations. It is a pleasure to thank the members of 3.6m DOT team and IR astronomy group at
TIFR for their support during TANSPEC observations. TIFR-ARIES Near Infrared Spectrometer (TANSPEC)
was built in collaboration with TIFR, ARIES and MKIR, Hawaii for the DOT. We thank the staff of IAO,
Hanle and CREST, Hosakote, that made these observations possible. The facilities at IAO and CREST are
operated by the Indian Institute of Astrophysics. JPN, DKO and KS acknowledge the support of the Department of Atomic Energy, Government of India, under project Identification No. RTI 4002.
We acknowledge with thanks the variable star
observations from the AAVSO International Database contributed by observers worldwide and used in this
research. We also acknowledge the ATLAS and the NEOWISE observations that have been used in this 
research. We also thank the BINA organisers for proving us with the opportunity to showcase our work 
in their platform.

\end{acknowledgments}

\vspace{-0.7cm}
\begin{furtherinformation}

\begin{orcids}
\orcid{0000-0001-7650-1870}{Arpan}{Ghosh}
\orcid{0000-0001-5731-3057}{Saurabh}{Sharma}
\orcid{2222-3333-4444-5555}{Joe}{Phillip Ninan}
\orcid{0000-0001-9312-3816}{Devendra}{K. Ojha}
%\orcid{0000-0002-6586-936X}{Aayushi Verma}
%\orcid{}{Tarak}{Chand Sahu}
%\orcid{}{Rakesh}{Pandey}
%\orcid{}{Koshvendra}{Singh}
\end{orcids}
%{\sl This section is optional.
%You may list here the ORCIDs of those authors who would like to share them, one per line, with the \verb|\orcid{|\texttt{\emph{ORCID}}\verb|}{|\texttt{\emph{First name}}\verb|}{|\texttt{\emph{Last %name}}\verb|}| command.
%This command typesets the information, and makes the ORCIDs themselves active links to the corresponding records on \href{https://orcid.org}{orcid.org}.

%Unlike in this sample, no other text should actually be included here and this section should reduce to a bare list.
%The \verb|\orcid| command controls line feeds by itself; please do not insert any \verb|\\| or \verb|%\newline| before or after them.}

\vspace{-0.7cm}
\begin{authorcontributions}

The lead author is responsible for the data acquisition, data analysis and paper writing while the project and methodology has been conceived by Saurabh Sharma, D. K. Ojha and J. P. Ninan. The rest of the co-authors have helped in developing the scientific interpretations that are presented in this
work.  

\end{authorcontributions}

\vspace{-0.5cm}
\begin{conflictsofinterest}
%This section is \emph{mandatory}.
%Authors must declare any personal or professional circumstances that may be perceived as influencing the research reported in the paper.
%If there is no conflict of interest, please state that ``
The authors declare no conflict of interest.
\end{conflictsofinterest}

\end{furtherinformation}

\bibliographystyle{bullsrsl-en}

\bibliography{bina}

\end{document}